\documentclass{aa}
\usepackage{graphics,epsfig,lscape}

\usepackage{natbib}
\bibpunct{(}{)}{;}{a}{}{,}

\newcommand{\ergcm}[1]{$\cdot 10^{#1}$ erg cm$^{-2}$ s$^{-1}$}

\newcommand{\hcm}[1]{$\cdot 10^{#1}$ cm$^{-2}$}
\newcommand{\ohcm}[1]{$10^{#1}$ cm$^{-2}$}

\newcommand{\kms}{km\,s$^{-1}$}
\newcommand{\ktbb}{kT$_{\rm bb}$}
\newcommand{\ktlmk}{kT$_{\rm MK}^{\rm s}$}
\newcommand{\kthmk}{kT$_{\rm MK}^{\rm h}$}
\newcommand{\eqw}{Fe$_{\rm eqw}^{6.4}$}
\newcommand{\nbb}{norm$_{\rm bb}$}
\newcommand{\nlmk}{norm$_{\rm MK}^{\rm s}$}
\newcommand{\nhmk}{norm$_{\rm MK}^{\rm h}$}
\newcommand{\nh}{\hbox{N$_{\rm H}$}}

\newcommand{\ct}{cts s$^{-1}$}

\newcommand{\Halp}{H${\alpha}$}
\newcommand{\HeIIX}{\ion{He}{II} $\lambda$ 4686}
\newcommand{\HeI}{\ion{He}{I}}
\newcommand{\HeII}{\ion{He}{II}}
\newcommand{\cn}{\ion{N}{III}-\ion{C}{III} ${\lambda \lambda}$4640-60}
\newcommand{\Hbet}{H${\beta}$}
\newcommand{\Hgam}{H${\gamma}$}

\newcommand{\FeXXV}{\ion{Fe}{XXV}}
\newcommand{\FeXXVI}{\ion{Fe}{XXVI}}
\newcommand{\rxj}{\hbox{\object{1RXS\,J154814.5-452845}}}
\newcommand{\vcas}{\hbox{\object{V\,709\,Cas}}}
\begin{document}
 
\title{X-ray and optical observations of \rxj: a new 
       intermediate polar with soft X-ray emission
       \thanks{Based on observations with XMM-Newton, an ESA Science Mission
               with instruments and contributions directly funded by ESA Member
               states and the USA (NASA) and on observations collected at the
	       European Southern Observatory}}
 
\author{F.~Haberl\inst{1} \and C.~Motch\inst{2} \and F.-J.~Zickgraf\inst{3}}

\titlerunning{\rxj, a new intermediate polar}
\authorrunning{Haberl et al.}
 
\offprints{F. Haberl}
 
\institute{Max-Planck-Institut f\"ur extraterrestrische Physik,
           Giessenbachstra{\ss}e, 85748 Garching, Germany
	   \and
	   Observatoire de Strasbourg, 11, rue de l'Universite,
           67000 Strasbourg, France
	   \and
	   Hamburger Sternwarte,
	   Gojenbergsweg 112, 21029 Hamburg, Germany}

\date{Received 21 December 2001/ Accepted 5 March 2002}
 
\abstract{
We report the identification of the ROSAT all-sky survey source \rxj\ as
new intermediate polar and present the results from follow-up optical and 
X-ray observations. The source shows pulsations with a period of 693 s
both in the optical and X-ray light curves and the detection of a synodic 
frequency strongly suggests that this is the rotation period of the white
dwarf. Although the one day aliasing and the sparse optical data 
coverage does not allow to unambiguously identify the orbital period,
the most likely values of 9.37 h and 6.72 h add \rxj\ to the intermediate 
polars with the longest orbital periods known. 
The optical spectrum displays features from the late type secondary and 
shows the presence of broad absorption lines at 
\Hbet\ and higher order Balmer lines which may be a signature of the 
white dwarf atmosphere, very similar to \vcas\
\citep[RX\,J0028.8+5917][]{2001A&A...374.1003B}. 
The average X-ray spectra as obtained by the EPIC instruments on board
XMM-Newton show hard emission typical for this class of objects but also
the presence of soft blackbody-like emission similar to that seen from soft intermediate 
polars and thought to arise from the white dwarf surface heated by the
hard X-rays. The best fit model comprises thermal emission from multi-temperature 
plasma in collisional ionization equilibrium with a 
continuous temperature distribution up to a maximum of $\sim$60 keV, 
an Fe fluorescence line at 6.4 keV and with equivalent 
width of 260 eV and a blackbody component with kT of 86 eV. The hard X-ray emission 
is absorbed by matter covering 47\% of the X-ray source with an equivalent
hydrogen density of $\sim$\ohcm{23}. The remaining hard emission is absorbed 
by a much reduced column density of 1.5\hcm{21} as is the soft blackbody emission.
Pulse-phase spectroscopy around spin maximum and minimum reveals that the flux
variations are mainly caused by a change in the temperature distribution with
higher intensity (a factor of $\sim$3 in the 1 keV emission) seen from the lower 
temperature plasma during spin maximum. The absorption in the high column 
density matter only decreases marginally during spin maximum. The emission
characteristics are consistent with the accretion curtain scenario and features 
in the X-ray pulse profiles indicate that we observe one pole of the white dwarf
and our line of sight is nearly parallel to the curtain at spin minimum while
at maximum we have a more direct view to the cooling post shock accretion flow.
\keywords{Binaries: close -- Stars: individual: \rxj\ -- 
          Stars: novae, cataclysmic variables -- X-rays: stars}}                                                                  
 
\maketitle
 
\section{Introduction}

The ROSAT Galactic Plane Survey project \citep[RGPS,][]{1991A&A...246L..24M}  
to identify X-ray sources detected at low galactic latitudes 
($\vert$b$\vert$ $<$ 20\degr) led to the discovery of new cataclysmic 
variables \citep{1995A&A...297L..37H,1996A&A...307..459M}. 
Cataclysmic variables (CVs) are close 
binary systems in which the low mass secondary star fills
its Roche lobe and transfers matter onto the white dwarf primary. 
Depending on the magnetic field strength of the white dwarf the accretion 
flows via an accretion disc or directly couples
onto the magnetic field lines of the white dwarf in the case of a strong
magnetic field. Magnetic cataclysmic variables are usually divided 
into polars (AM Her systems) in which the strong magnetic field 
forces the two stars into co-rotation and intermediate polars with shorter 
white dwarf spin periods of typically a few hundred seconds
\citep[IPs, for reviews of this latter sub-class see][]
{1994PASP..106..209P,1995mcv..conf..185H,1999mcv..work....1H}.
IPs accrete from the accretion disc via an azimuthally extended
accretion curtain which causes high photo-electric absorption.

The X-ray spectra of most intermediate polars show a hard,
absorbed bremsstrahlung continuum plus emission lines from highly ionized
iron. This emission originates from the matter which free-falls towards
the white dwarf creating a strong shock near the surface with temperatures
kT up to several tens of keV. The hard X-rays heat the white dwarf surface, 
creating a soft blackbody-like component as is generally observed in polars 
\citep{1993AdSpR..13..115B} where the accretion stream is of pencil-like 
geometry. Additional heating may be caused by dense blobs in the accretion
flow which directly penetrate the white dwarf atmosphere. 
The temperature of the emission is thought to decrease with increasing size of
the accretion region on the white dwarf \citep{1990MNRAS.247..214K}. 
Therefore, the absence of detectable soft X-ray emission in classical IPs 
may be explained by a too low temperature and/or suppression of soft X-rays
due to considerable absorption. The discovery of little absorbed, 
soft X-ray emission from a small number of `soft' intermediate polars in 
ROSAT data \citep{1992MNRAS.258..749M,1994A&A...291..171H} together with 
relatively high magnetic field strengths \citep[e.g.][]{1997MNRAS.285...82P}
which collimates the accretion stream is consistent with this picture.
Here we present the discovery of a new cataclysmic variable. Optical
observations together with detailed X-ray follow-up using ROSAT and XMM-Newton
clearly identify \rxj\ as new intermediate polar. 

\section{Discovery in the ROSAT all-sky survey}

ROSAT detected \rxj\ during the all-sky survey \citep{1982AdSpR...2..241T} 
with 0.54$\pm$0.06 \ct\ in the PSPC detector \citep{1987SPIE..733..519P}. 
The source is listed in the bright source catalogue \citep{1999A&A...349..389V}
with hardness ratios (defined as
HR1 = (H--S)/(S+H) and HR2 = (H2--H1)/(H1+H2) where S, H, H1 and H2 
denote count rates in the 0.1--0.4 keV, 0.5--2.0 keV, 0.5--0.9 keV 
and 0.9--2.0 keV bands, respectively) which indicate a moderately hard 
X-ray spectrum in the ROSAT band. Table~\ref{ip-hr} compares the hardness 
ratios of \rxj\ to four other ROSAT-discovered IPs extracted from the 
RRA PSPC source catalogue from
pointed ROSAT observations \citep{2000rra}. Classical IPs like RX\,J0028.8+5917 and 
RX\,J1712.6--2414 show high absorption, for which HR1 is very sensitive to
with values approaching +1. The `soft' IPs RE\,0751+14 and RX\,J0558.0+5353 
show little absorption (negative HR1) but also softer intrinsic spectra
(negative HR2) caused by a strong soft blackbody spectral component which
dominates the spectrum in the ROSAT band. The hardness ratios of \rxj\ 
are between those of the two groups, suggesting less absorption than seen 
from classical IPs and/or the existence of a soft spectral component. 
\begin{table}
\caption[]{ROSAT hardness ratios of IPs}
\begin{tabular}{lcc}
\hline\noalign{\smallskip}
\multicolumn{1}{l}{IP} &
\multicolumn{1}{c}{HR1} &
\multicolumn{1}{c}{HR2} \\ 
\noalign{\smallskip}\hline\noalign{\smallskip}
 \rxj              & +0.80$\pm$0.07  & +0.06$\pm$0.11 \\
\noalign{\smallskip}
 RX\,J0028.8+5917  & +0.95$\pm$0.01  & +0.37$\pm$0.01 \\
 RX\,J1712.6--2414 & +0.92$\pm$0.01  & +0.32$\pm$0.01 \\
 RE\,0751+14$^1$   & --0.83/--0.79   & --0.09/--0.15 \\
 RX\,J0558.0+5353  & --0.45$\pm$0.01 & --0.09$\pm$0.02 \\
\noalign{\smallskip}
\hline
\end{tabular}

$^1$ observed twice with slightly different hardness ratios
\label{ip-hr}
\end{table}

\section{Optical observations}

\subsection{Optical identification}

\begin{figure}
\resizebox{\hsize}{!}{\includegraphics[bb=50 55 570 625,angle=-90,clip]{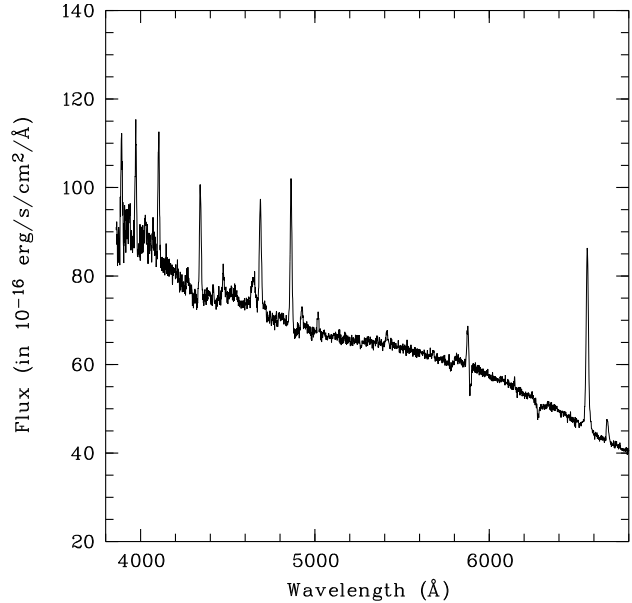}}
\caption{Mean optical spectrum of the optical counterpart of \rxj\ as observed in June 1997}
\label{97spec}
\end{figure}

\begin{figure}
\resizebox{\hsize}{!}{\includegraphics[]{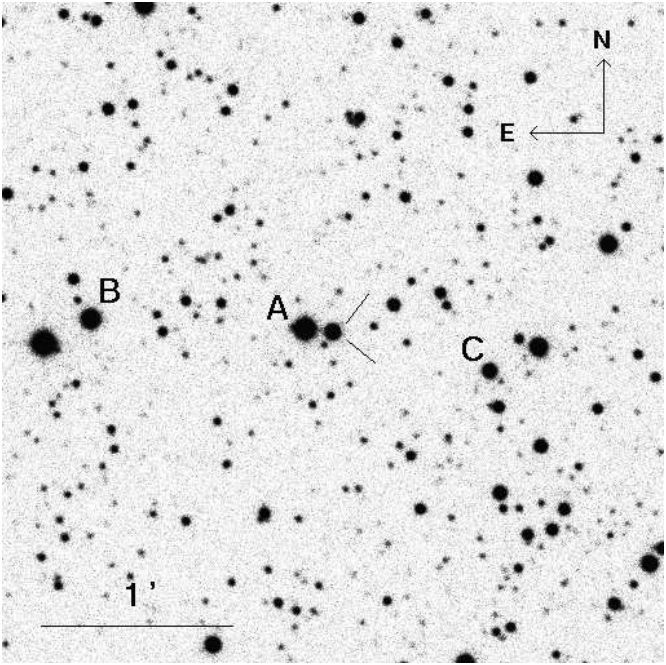}}
\caption{V band finding chart for \rxj. The positions of the three comparison stars, A,B and C 
         used for differential photometry are also shown}
\label{fc1548}
\end{figure}

The source was included in the RGPS bright source sample for optical identification. We observed
the field of \rxj\ for the first time on 1997, June 6 with DFOSC and the LORAL 2k x 2k CCD at the
ESO-Danish 1.54m telescope. The X-ray source was readily identified with a V $\sim$ 14.6 blue
object exhibiting Balmer, \HeI , \HeII\ and \cn \ lines in emission. We show in Fig.~\ref{97spec} 
the mean of two 10\,mn long spectra obtained with a 2.5 \arcsec \ slit through grism \#7 ($\lambda
\lambda$ 3860 - 6810 \AA ; 10 \AA \ FWHM resolution). Fig.~\ref{fc1548} shows the position of the
optical counterpart on a V band image.

\subsection{Follow-up optical observations}

We organized an optical observation campaign in May 1998 involving both photometric and
spectroscopic monitoring at the ESO-Dutch 0.9m and ESO-Danish 1.54m telescopes, respectively.
During three consecutive nights from May 7 to 9 UT we obtained optical V band photometry using
the standard CCD camera equipped with the TK512CB \# 33 CCD. In order to ensure the best
compromise between time resolution and observing efficiency the CCD was windowed and the exposure
time was set to 30\,s. With this instrumental setting, the mean time between two consecutive
exposures was $\sim$ 53\,s. A total of 753 frames were acquired over 11.5\,h. Raw images were
corrected for bias and flat-field using standard MIDAS procedures. Instrumental magnitudes were
computed using the SExtractor package \citep{1996A&AS..117..393B} 
for the optical counterpart and 3
comparison stars, named A, B and C whose positions are indicated in Fig. \ref{fc1548}. Weather
conditions were not photometric during the first and the third night. However, the accuracy of
the differential photometry remains good with average errors below the 1\% level. From standard
star measurements in June 1997 and May 1998 we derive the following magnitudes for the
comparison stars; V$_{\rm A}$ = 12.94 $\pm$ 0.04, V$_{\rm B}$ = 13.51 $\pm$ 0.04 and V$_{\rm C}$
= 15.00 $\pm$ 0.04. The photometric uncertainty is dominated by the colour transformation. The
mean magnitude of \rxj\ recorded in May 1998 was V = 14.61, somewhat fainter than in June 1997,
V = 14.46. Spectroscopic observations took place from May 15 till 17 UT. At the ESO-Danish 1.54m
telescope, we used the same DFOSC setting as in 1997 but with a smaller entrance slit of 1.5
\arcsec . With grism \#7 the wavelength range was $\lambda \lambda$ 3860 - 6840 \AA\ with a 5.7 \AA
\ FWHM resolution). Because of the smaller slit used in 1998 than in 1997, the spectral
resolution improved but no accurate flux calibration could be established. In particular,
differential refraction distorted the shape of the flux calibrated spectrum. The first spectrum
had an exposure time of 20\,mn and the remaining 8 were 15\,mn long. Arcs from He and Ne lamps
provided the wavelength scale. As for photometric data, MIDAS procedures were used to correct 2-d
frames and extract and calibrate spectra in wavelength and flux.

\subsection{Spectroscopic analysis}

\subsubsection{The orbital period}

\begin{figure}
\resizebox{\hsize}{!}{\includegraphics[bb=50 55 570 625,angle=-90,clip]{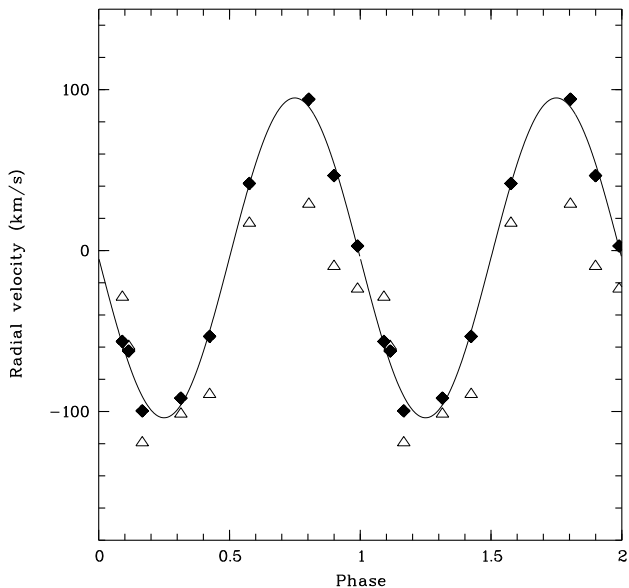}}
\caption{Example of radial velocity curves: \Halp \ velocities (filled diamond) 
         folded with the best fit sine of period P = 9.65 $\pm$ 0.03\, h. Other 
	 periods close to 6.7 \,h are also consistent with the spectroscopic data. 
         \HeIIX \ radial velocities shown as open triangles follow a less regular 
	 pattern of variation}
\label{plotrv2}
\end{figure}

Radial velocities of Balmer and of \HeIIX \ emission lines vary relatively slowly within a single
night. The total amplitude is 200 \kms \ for \Halp . Balmer and \HeII \ velocities seem roughly in
phase although there are large differences in mean velocities, higher Balmer lines being more blue
shifted than \Halp . We list in Table \ref{rvtab} the measured emission line velocities. In order
to search for orbital motion we computed periodograms for the various lines using the method
outlined in \citet{1976Ap&SS..39..447L} 
and \citet{1982ApJ...263..835S} 
which is suitable for unequally spaced data.
Spectroscopic data are consistent with a rather long orbital period, larger than 5\,h . Because of
the strong one day aliasing and sparse coverage it is not possible to unambiguously derive the
orbital period. We find that the two most probable periods are  P = 9.37 $\pm$ 0.69\,h and P = 6.72
$\pm$ 0.32\,h. We show in Fig. \ref{plotrv2} the \Halp \ radial velocity curve folded with a
period of 9.65 $\pm$ 0.03\, h which gives the best sine fit. At this period, \HeIIX \ velocities
have a more structured pattern of variation. However, the overall shape of the radial velocity
curve changes with period and clearly more spectroscopic data are needed to determine the true
orbital period and velocity structure. 

A rather strong indication in favour of a long orbital period is the presence of several metallic
absorption lines in the red part of the spectrum. We compare in Fig.~4 
the red part of
the mean optical spectrum obtained in 1998 with a template K4 III spectrum extracted from the
\citet{1984ApJS...56..257J} 
library of stellar spectra. The resolution of the library spectra
($\sim$ 4.5 \AA) compares well with that of our observations. In order to better highlight the
similarities we diluted the late type spectrum by a factor 5 using a constant continuum and plotted
together the rectified spectra normalized to the average continuum. The Mg "b" bands and the
$\lambda$ 5206, 5270, 6362, 6497 \AA \ atomic line blends are visible as well as the \ion{Ca}{I}
$\lambda$ 6162 \AA \ line. The Na I doublet is well marked, however, part of its strength could be
due to interstellar absorption as indicated by the presence of a relatively deep $\lambda$ 5780
\AA\ band. The large absorption at $\lambda$ 6280 \AA\ is probably a mixture of telluric O2 and
diffuse interstellar bands. It is not possible to classify the late type spectrum with great
accuracy. \citet{1993PASP..105..693T} 
have identified useful lines for classifying stars in the
near-infrared.  The absence of TiO bands at $\lambda$ 6080 - 6390 \AA\ indicates a spectrum earlier
than  K7V. The marked $\lambda$ 6497 \AA\ blend and \ion{Ca}{I} line suggests an early K type.
Finally, the presence of the \ion{Fe}{I}, \ion{Ti}{I} \ion{Cr}{I} blend at $\lambda$ 6362 \AA\ may
indicate an evolved type. Unfortunately, the absorption line depths on individual 15\,mn long
spectra are not strong enough to measure a reliable radial velocity curve  of the late type
companion. 

\begin{table}
\caption{Radial velocities corrected at barycentre of solar system in \kms.}
\label{rvtab}
\begin{tabular}{rrrrr}
        Julian date & \Halp & \Hbet & \Hgam & \HeII \\
     -2,400,000.5 d &       &       &       &  $\lambda$ 4686      \\ \hline
        50948.28693 & +94.1 & +33.8 & +12.0 & +28.9 \\
        50948.32557 & +46.6 & -30.2 & -34.1 &  -9.8 \\
        50948.36179 &  +2.9 & -61.5 & -69.5 & -23.9 \\
        50948.41196 & -62.4 &-102.8 &-121.1 & -59.5 \\
        50949.23678 & -99.6 &-141.2 &-167.1 &-119.4 \\
        50949.29622 & -91.7 &-120.1 &-132.0 &-101.5 \\
        50949.34037 & -53.3 & -45.7 & -52.4 & -89.3 \\
        50949.40147 & +41.7 & +43.8 & +11.3 & +17.1 \\
        50950.41219 & -56.5  &-69.7 & -60.2 & -28.9 \\
\hline
\end{tabular}
\end{table}

A spectral type hotter than K7 and possibly early K is compatible with the long orbital period
derived from our spectroscopy. The mean density of the Roche lobe filling star can be expressed as
$\rho \approx 115$\,P$_{\rm hr}^{-2}$\,g\,cm$^{-3}$ 
\citep{1992apa..book.....F}. 
The shortest
possible period of 6.72\,h implies densities comparable to those of a main sequence K5 star. If the
orbital period is 9.37\,h a main sequence star would have a G type. Although this cannot be ruled
out completely from our data, the requirement for stable mass transfer of 
q = M$_{2}$/M$_{1}$ $\leq 5/6$ \citep{1992apa..book.....F} 
would rather suggest a slightly evolved K secondary instead.
This picture is consistent with the spectral type distribution of secondary stars in cataclysmic
variables compiled by \citet{1998MNRAS.301..767S} 
which shows that no system below 4.3\,h period has a secondary earlier than K. 

Standard evolutionary sequences for unevolved donor stars undergoing
mass-transfer predict a $\sim$ K2 spectral type if the orbital period of
the CV is 6.72h \citep{2000MNRAS.318..354B}. 
Alternatively, a K type secondary
in a 9.37h period CV would only be possible if the mass donor star is
slightly evolved \citep{1998A&A...339..518B}. 
Therefore,
in the absence of confirmed orbital period and secondary mass estimates,
the evolutionary status of the mass donor star cannot be determined. 

\begin{figure*}
\resizebox{12cm}{!}{\includegraphics[]{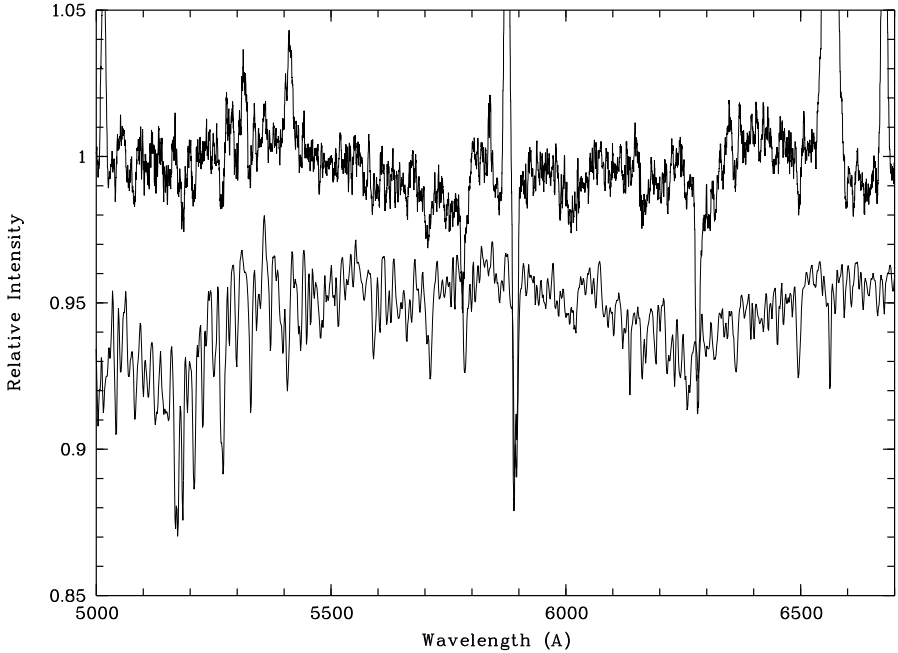}}
\hfill
\parbox[b]{55mm}{
\caption{The red part of the averaged spectrum of \rxj\ as observed in 1998 (top) 
         plotted together with a template stellar spectrum of a K4 III star diluted 
	 with a constant continuum. Several faint metallic lines visible in the late 
	 type spectrum also appear in that of \rxj, revealing the late type secondary 
	 (see text)}}
\label{latetype}
\end{figure*}

\subsubsection{Spectral signature of the white dwarf ?}

\begin{figure}
\resizebox{\hsize}{!}{\includegraphics[bb=50 55 570 625,angle=-90,clip]{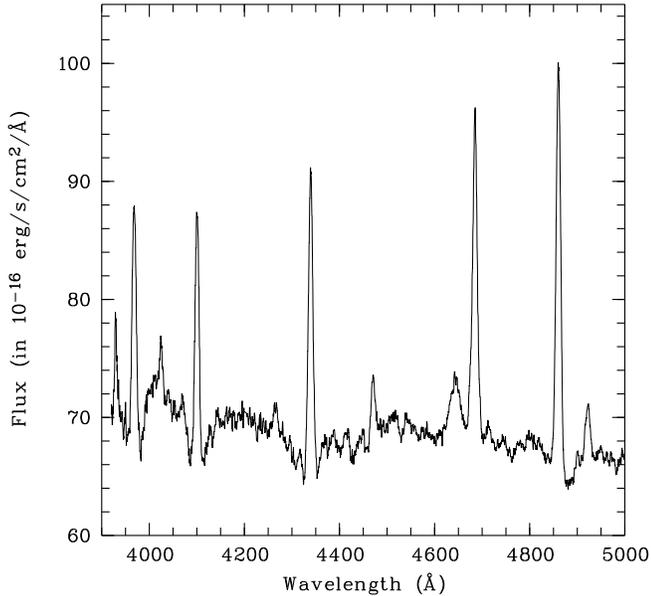}}
\caption{Blue part of the average 1998 spectrum of \rxj. The broad absorption 
         features underneath the emission spectrum could be the signature of the 
	 white dwarf photosphere}
\label{98spec}
\end{figure}

Another outstanding feature of \rxj\ is the presence of broad absorption lines at \Hbet\ and
higher order Balmer lines. Troughs are seen up to H$\epsilon$ in the blue part of the spectrum
(see Fig. \ref{98spec}). In that respect, the spectrum of \rxj\ is strikingly similar to that of
the ROSAT discovered classical intermediate polar \vcas\
\citep[RX\,J0028.8+5917][]{2001A&A...374.1003B}. 
These authors argue that the broad absorptions are the signature of the non heated part of the
accreting star and derive log g = 8 and T = 23,000 K for the DA white dwarf. Similar conclusions
may be driven for \rxj. One should however take this  interpretation with some caution since
similar broad Balmer lines in absorption with superposed narrow emissions are sometimes seen from
accretion discs in UX UMa stars \citep[see e.g.][ and references therein]{1995cvs..book.....W}. 

\subsection{Photometric timing analysis}

We show in Fig.~6 
the V light curve of \rxj\ for the three consecutive nights of
observations. Significant photometric activity occurs all along the light curve on time
scales of minutes. No conspicuous large long term variations indicating a possible orbital modulation is seen
although the hollow at the beginning of the second night may represent some minimum. In order to
search for coherent modulations we computed the Lomb-Scargle periodogram of the entire time series.
Fig.~\ref{photoper} shows the resulting power spectrum. A clear signal is detected at $f$ = 124.72
$\pm$ 0.10 cycle/day or P = 692.75 $\pm$ 0.56\,s. This period is consistent with the most probable
X-ray period derived from the long HRI observation (P = 693.3 $\pm$ 1.1\,s, see below). 
Although the two closest one day aliases of the photometric period cannot strictly be ruled 
out on a statistical basis, none of them appears consistent with any of the possible HRI aliases 
(see Fig.~\ref{photoperdetails}). The natural aliasing frequencies of ground based and space borne
instruments are obviously different. It is therefore very likely that the optical pulsation
occurs at the same frequency as in X-ray and probably represents the spin period of the accreting
white dwarf unless the accretion is disc-less as supposed in RX\,J1712.6-2414 for instance 
\citep{1997MNRAS.287..117B}. 
The absence of large power at harmonic frequencies is consistent with the sinusoidal
shape of the folded light curve shown in Fig. \ref{foldphot}. The full amplitude of the folded V
light curve is 0.036 mag. well in the range of that observed in other intermediate polars 
\citep[see e.g.][]{1994PASP..106..209P,1996MNRAS.282..739W}.

We investigated further the presence of beat frequencies by applying the Lomb-Scargle periodogram
to data with the main modulation at  P = 692.75\,s removed. The resulting periodogram, plotted in
Fig.~\ref{photoperdemod}, exhibits an excess power in the interval 120 to 125 cycle/day. The
highest peak is at 122.14 $\pm$ 0.14 d$^{-1}$ (P = 707.39 $\pm$ 0.21\,s). At  this given
frequency, the formal probability of the peak to be real is 1.7\% 
assuming a white noise local distribution. The value of the peak
frequency agrees very well with that expected for the synodic frequency ($f$ = 122.14 $\pm$ 0.21
d$^{-1}$) computed assuming the best orbital period of P = 9.37 $\pm$ 0.69\,h, derived from the
optical spectroscopy, and a white dwarf spin period of P = 692.75 $\pm$ 0.56\,s, derived from the
optical photometry. However, an orbital period of $\sim$ 6.7\,h cannot be strictly ruled out by
the observations. At the synodic period the light curve is nearly sinusoidal and has a full
amplitude of 0.023 mag. The detection of a synodic frequency confirms that the main X-ray and
optical period is the true rotation period of the white dwarf.

\begin{figure*}
\resizebox{12cm}{!}{\includegraphics[]{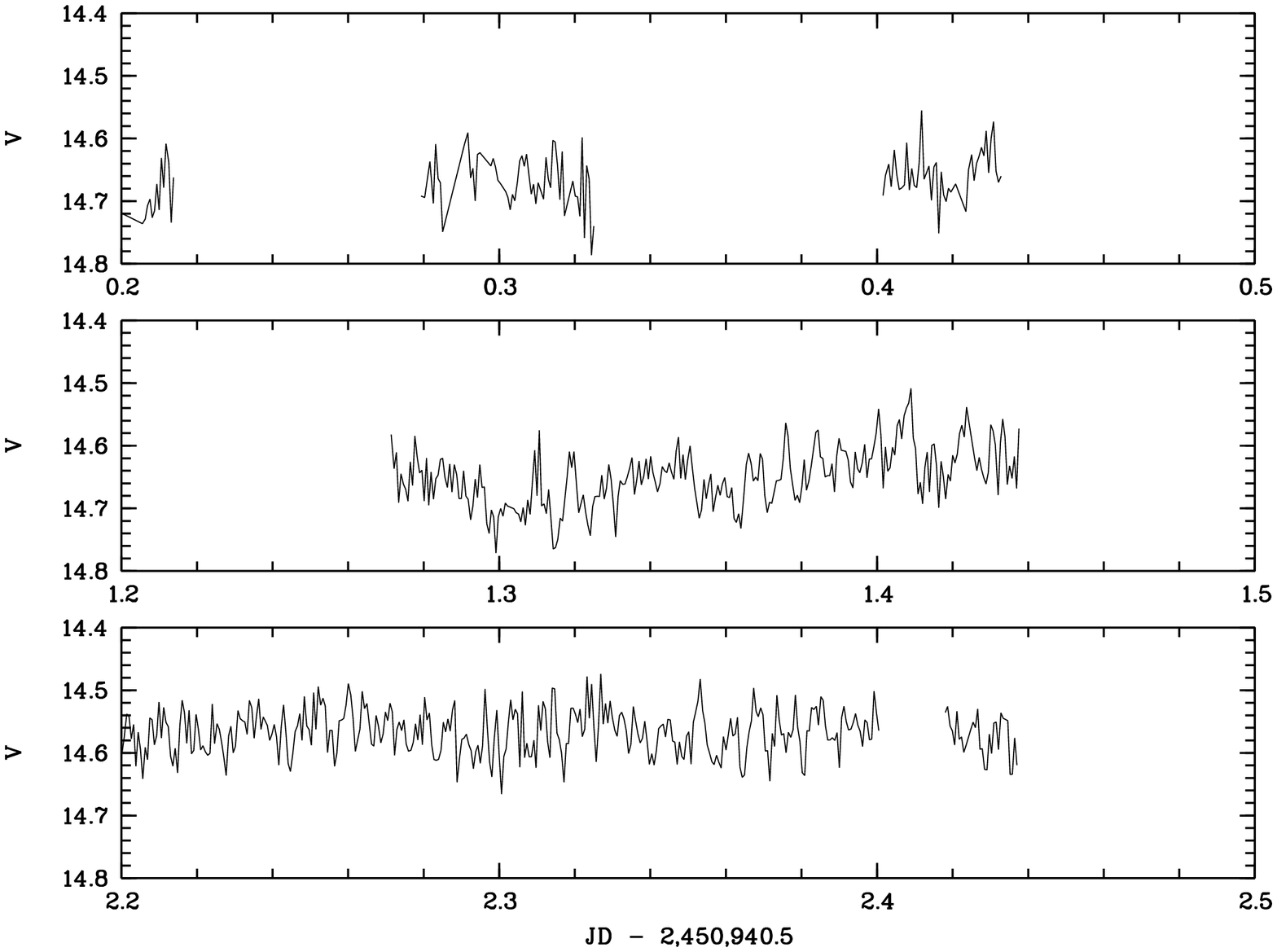}}
\hfill
\parbox[b]{55mm}{
\caption{Optical V band light curve of \rxj\ obtained in May 1998 with the ESO-Dutch 
         0.9m telescope}}
\label{plotphoto1}
\end{figure*}

\section{ROSAT soft X-ray observations}

We proposed two observations of \rxj\ in the guest observer program of 
ROSAT using the HRI detector \citep{dhk96}. During the first short observation 
in 1996, Sept. 13 (RH400792, net exposure time of 1.97 ks) the average count rate was 
$0.22 \pm 0.01$ \ct. It was dedicated to derive an accurate X-ray position:
RA = $15^{\rm h}48^{\rm m}14.5^{\rm s}$, Dec = $-45\degr28\arcmin39\arcsec$ with
an error of $\sim7\arcsec$ dominated by systematic boresight uncertainty.
After the optical identification as cataclysmic variable and IP candidate
a second ROSAT HRI observation (RH300624) was performed in 1998, Sept. 11-14, with a net
exposure of 17.89 ks. The flux with an average count rate of $0.190\pm0.006$ \ct\ 
was similar to the first HRI observation and also to the survey detection
(the HRI to PSPC count rate conversion factor is 3.0 for a hard spectrum).

The second HRI observation, performed near the end of the ROSAT mission, 
suffered from unstable attitude and no accurate position could be derived. 
However, the data can be used for a temporal analysis. The HRI light 
curve (0.1--2.4 keV) shows sinusoidal intensity variations by a factor 
of about two. A Z$^2$ test \citep{1983A&A...128..245B} yields formally
a period of 693.3$\pm$1.1 s consistent with the value derived from the 
optical observations and 
confirming \rxj\ as IP, however, the large number of short observation 
intervals typically lasting less than 2000 s causes many alias peaks. 
Periods of 682.2 s, 627.5 s and 618.4 s are of statistically similar 
significance, but are inconsistent with any of the possible alias periods
seen in the optical (Fig.~\ref{hri-ztest}).
The folded soft X-ray light curve obtained from the HRI data using the 693.3 s
period is presented in Fig.~\ref{hri-efold}.

\begin{figure}
\resizebox{\hsize}{!}{\includegraphics[bb=50 55 570 625,angle=-90,clip]{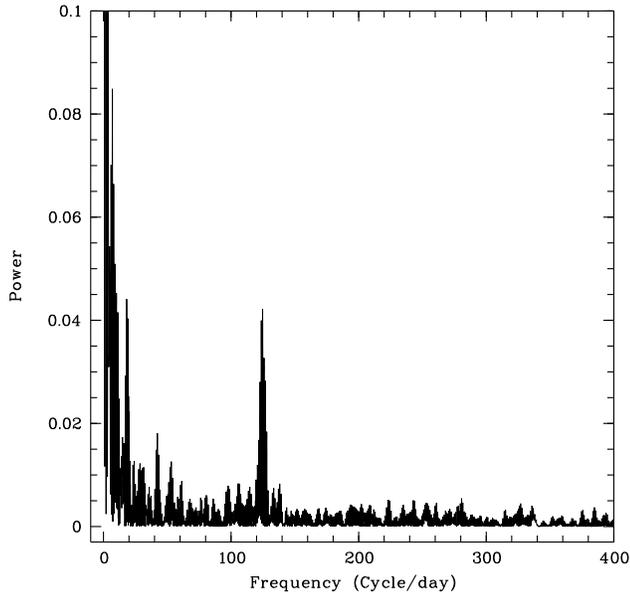}}
\caption{Periodogram of the V band time series obtained in May 1998 at the 
         ESO-Dutch telescope. A clear periodic signal at $f$ = 124.72 $\pm$ 0.10 
	 cycle/day (P = 692.75 $\pm$ 0.56 s) is detected.
         This period is consistent with that seen in the HRI observation}
\label{photoper}
\end{figure}

\begin{figure}
\resizebox{\hsize}{!}{\includegraphics[bb=50 55 570 625,angle=-90,clip]{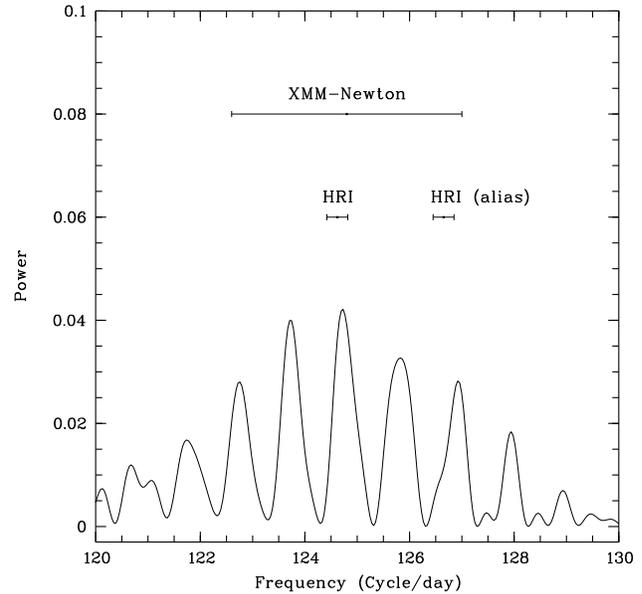}}
\caption{Detail of the periodogram of the V band time series obtained in May 1998 
         at the ESO-Dutch telescope. The best optical photometric frequency is 
	 consistent with the most likely HRI determination} 
\label{photoperdetails} 
\end{figure}

\begin{figure}
\resizebox{\hsize}{!}{\includegraphics[bb=50 55 570 625,angle=-90,clip]{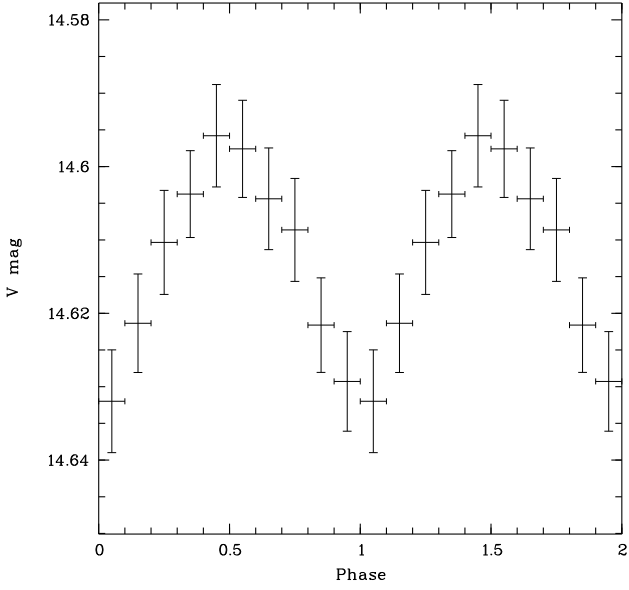}}
\caption{The V band light curve obtained in 1998 folded with the best period of 
         P = 692.75\,s. The mean full amplitude of the modulation is 0.036 mag}
\label{foldphot}
\end{figure}

\begin{figure}
\resizebox{\hsize}{!}{\includegraphics[bb=50 55 570 625,angle=-90,clip]{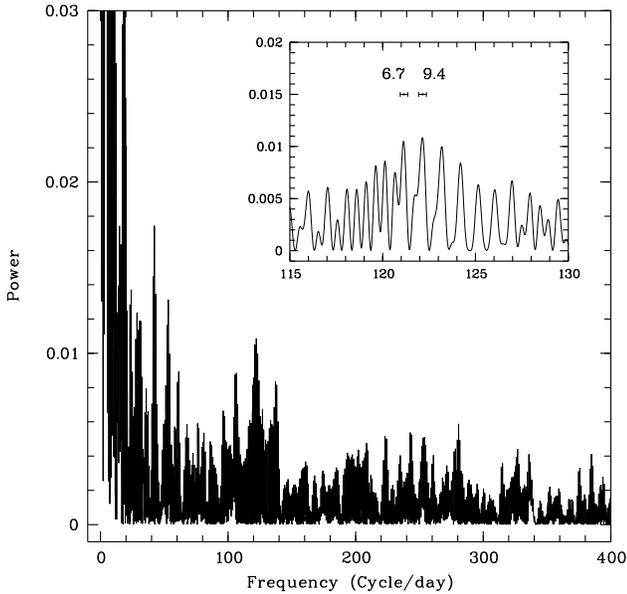}}
\caption{Periodogram of the 1998 V band time series demodulated from the main
         signal at P = 692.75 $\pm$ 0.56 s. Excess power is seen at the expected 
         position of the synodic frequency. The insert shows were the synodic 
	 frequencies should be assuming different orbital periods and the best 
	 optically determined white dwarf rotation period. The highest peak 
	 of the periodogram corresponds to the best spectroscopic candidate period 
	 of 9.4\,h}
\label{photoperdemod}
\end{figure}

\begin{figure}
\resizebox{\hsize}{!}{\includegraphics[angle=-90,clip,bb=78 40 570 700]{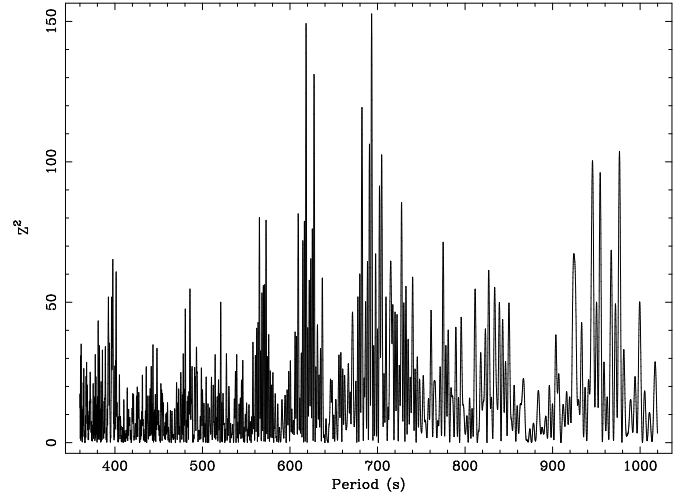}}
\caption{Z$^2$ test of the ROSAT HRI data
        }
\label{hri-ztest}
\end{figure}
\begin{figure}
\resizebox{\hsize}{!}{\includegraphics[angle=-90,clip,bb=77 38 570 700]{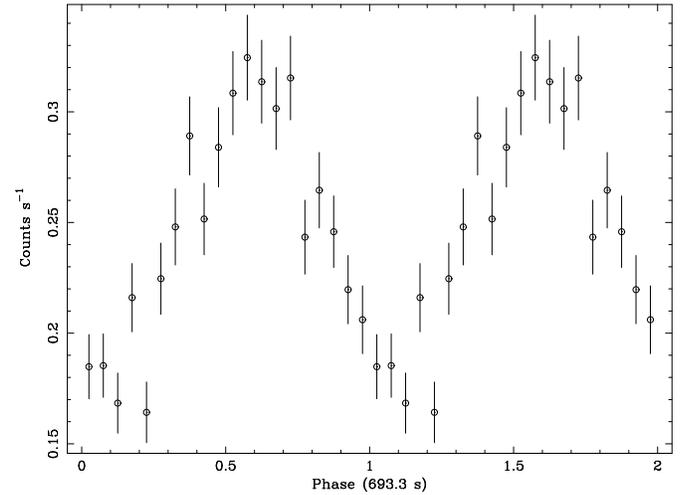}}
\caption{ROSAT HRI light curve of \rxj\ in the 0.1 -- 2.4 keV energy band
folded with a period of 693.3 s
        }
\label{hri-efold}
\end{figure}

\section{XMM-Newton broad band X-ray observations}

XMM-Newton \citep{2001A&A...365L...1J} observed \rxj\ as part of the 
telescope scientist guaranteed 
time during satellite revolution 137. The observation (id 0105460301)
on 2000, Sept. 7 lasted for about 20 ks and included the EPIC instruments.
The Reflection Grating Systems were switched off due to technical 
problems and the Optical Monitor was not used due to a bright star in the 
field. Here we report on the results from the European Photon Imaging 
Cameras based on MOS \citep[EPIC-MOS1 and -MOS2,][]{2001A&A...365L..27T} 
and PN CCD X-ray detectors \citep[EPIC-PN,][]{2001A&A...365L..18S} which 
are mounted behind the three telescopes \citep{2000SPIE.4012..731A}. 
The background in the EPIC detectors was low
during the whole observation with the exception of a short interval of 
$\sim$400 s when it increased from 1-2 \ct\ to about 10 \ct\ per PN CCD.
Source and background events were extracted from circular regions with radius
40\arcsec\ around the source position and a nearby source-free area in 
the image. 

\subsection{Temporal behaviour}

\begin{figure}
\resizebox{\hsize}{!}{\includegraphics[angle=-90]{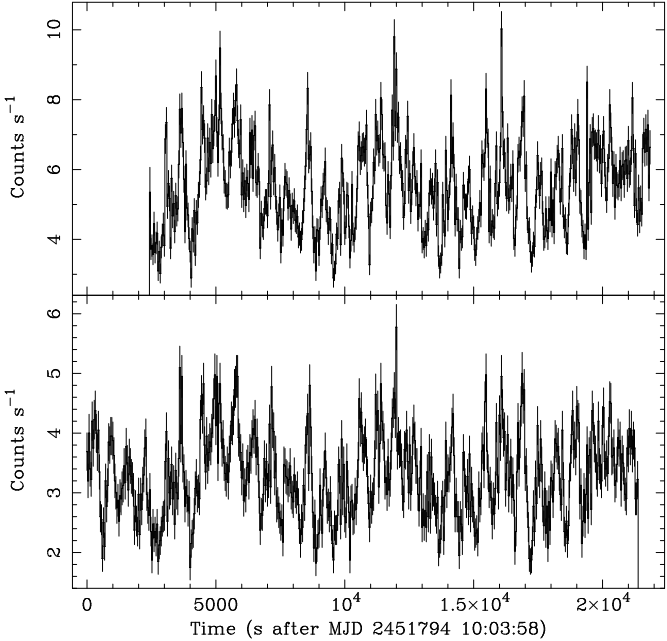}}
\caption{X-ray broad-band (0.2 -- 8.0 keV) light curves of \rxj\ obtained from the XMM-Newton 
         EPIC PN (top) and summed MOS1 and MOS2 (bottom) data with a  
         time binning of 40 s. The average background contribution is 
	 $3.6 \cdot 10^{-2}$ \ct\ for PN and $3.3 \cdot 10^{-2}$ \ct\ for MOS1+2}
\label{epic-lc}
\end{figure}

\begin{figure}
\resizebox{\hsize}{!}{\includegraphics[angle=-90,clip,bb=62 38 530 700]{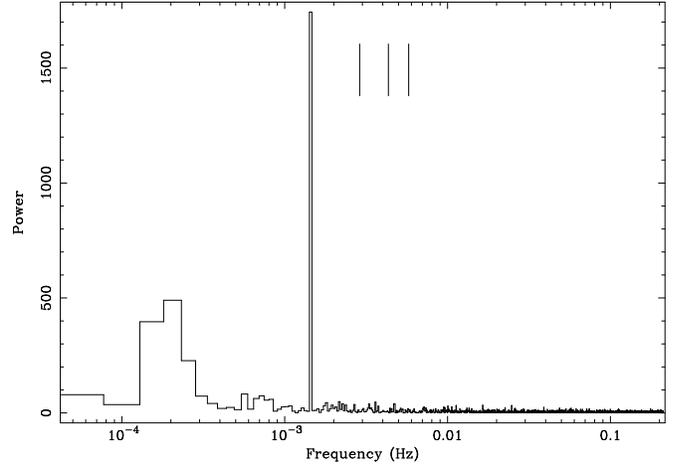}}
\caption{Power spectrum derived from of the XMM-Newton EPIC-PN data (0.2 -- 8.0 keV).
         The vertical lines indicate the frequencies of first, second and third harmonics
	 which are not detectable in the power spectrum}
\label{pn-powspec}
\end{figure}

\begin{figure}
\resizebox{\hsize}{!}{\includegraphics[angle=-90,clip,bb=62 38 556 700]{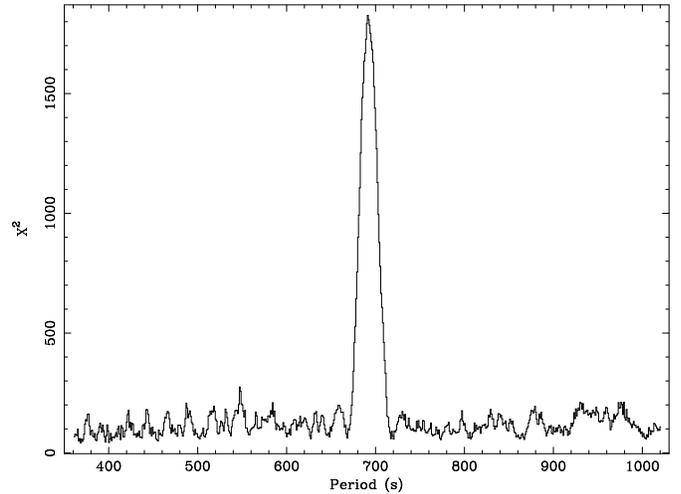}}
\caption{$\chi^2$ periodogram obtained from the XMM-Newton EPIC-PN data (0.2 -- 8.0 keV)
        }
\label{pn-chitest}
\end{figure}

The light curves in the 0.2--8.0 keV band obtained from PN and combined 
MOS data is plotted in Fig.~\ref{epic-lc}. Clearly seen are the periodic 
$\sim$700 s intensity variations on top of a more gradual change on a 
time scale of about 7000 s. A timing analysis using a Fast Fourier 
Transform (FFT) technique shows a single peak at the spin frequency 
with no detectable peaks at harmonic or sideband frequencies originating 
from the beat between spin and binary orbital period. This is consistent 
with a sinusoidal variation as is seen in the ROSAT HRI data
and a long orbital period in excess of the duration of the 
observation of $\sim$20 ks as is suggested by the optical data. The FFT power 
spectrum is shown in Fig.~\ref{pn-powspec}. There is some 
signature for a modulation on a longer time scale of 4000--8000 s in the 
power spectrum as was already seen in the light curve. Such a 
variation might be caused by the binary orbit, but the optical data are more
consistent with a longer orbital period.
A $\chi^2$ folding analysis was used to derive the best value of 692.3$\pm$12 s 
for the spin period. For comparison with Fig.~\ref{hri-ztest} the periodogram is
shown in Fig.~\ref{pn-chitest}. The contiguous XMM-Newton observation leads to a 
unique identification of the spin period, however, the shorter time base of the 
observation causes a larger uncertainty.


\begin{figure}
\resizebox{\hsize}{!}{\includegraphics[clip,bb=40 40 532 700]{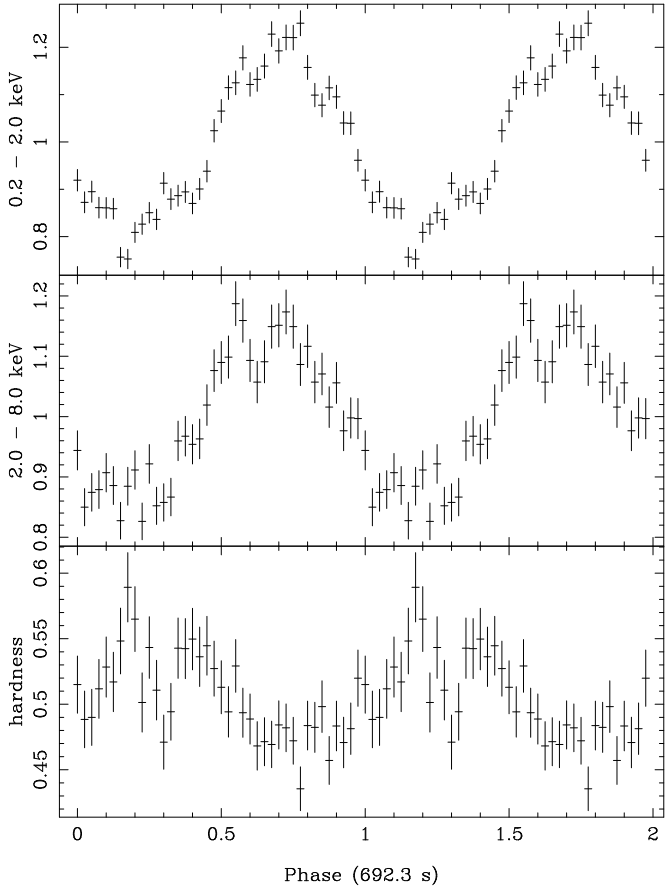}}
\caption{Folded light curves of \rxj\ derived from XMM EPIC-PN data in two energy bands
         together with their hardness ratio (2.0 -- 8.0 keV / 0.2 -- 2.0 keV). The folded light curves are
	 normalized to the average count rate (3.67 \ct\ in the 0.2 -- 2.0 keV band and 1.84 \ct\ in the 
	 2.0 -- 8.0 keV band and are plotted versus an arbitrary phase
        }
\label{pn-efold}
\end{figure}

The EPIC-PN data were folded with a period of 692.3 s in different energy bands 
(Fig.~\ref{pn-efold}). The 0.2--2.0 keV profile is consistent with that obtained from the 
HRI data (0.1--2.4 keV). The smaller statistical errors in the PN light curve allow to
identify two narrow features. A dip before the pulse maximum (or peak on the increasing flank) and
an asymmetric minimum are significant. The intensity dip is more prominent at energies 
2.0--8.0 keV while the minimum is more symmetric at high energies. The hardness ratio
(ratio of the count rates in the 2.0--8.0 keV to the 0.2--2.0 keV band) shows an increase 
during intensity minimum. The asymmetry in the soft light curve minimum reflects in a deep dip
in the hardness ratio, reaching the low level at intensity maximum. The intensity dip is 
apparently energy independent and not visible in the hardness ratio. The behaviour
of increased hardness during intensity minimum is often observed from IPs and is 
generally explained in the accretion curtain scenario by increased absorption 
during pulse minimum caused by the 
curtain. The sharp and short ($<$0.2 in phase) decline in hardness ratio near intensity 
minimum then suggests that the line of sight at this phase is nearly parallel to the curtain,
allowing an almost free view down to the white dwarf surface. As the white dwarf rotates
the curtain quickly intersects the line of sight causing increased absorption. 
The hardness ratio dip, the sinusoidal pulse variation with a single peak and the narrow 
optical emission lines suggest that we see emission from one pole only (if we indeed look 
down nearly parallel to the curtain the opposite pole in a dipole field should be hidden 
by the white dwarf at this spin phase). One should mention here that EX\,Hya also 
shows sinusoidal pulse variations while we see emission from both poles as indicated by, 
however, broad optical emission lines \citep{1987MNRAS.228..463H}.  

\subsection{The spin-phase averaged energy spectrum}

To further investigate the emission properties of \rxj\ spin-averaged energy 
spectra from the EPIC detectors were accumulated. From the PN data 
single- (pattern 0) and double-pixel (pattern 1-4) events were selected 
into two separate spectra to utilize the better energy resolution for 
single-pixel events. For the two MOS spectra pattern 0-12 events were used.
The four spectra were simultaneously fit with the same 
spectral model only allowing different normalizations between the two MOS 
and the PN spectra to account for possible inter-calibration problems in 
absolute fluxes and the different temporal coverage of the instruments 
(Fig.~\ref{epic-lc}). Spectral response files provided by the hardware
teams were used (epn\_ff20\_sY9\_medium.rmf and epn\_ff20\_dY9\_medium.rmf for PN
single and double events released in March 2001; m1\_medv9q19t5r4\_all\_15.rsp
and m2\_medv9q19t5r4\_all\_15.rsp for MOS released in Febr. 2001). In the past
various spectral emission and absorption components were required to fit X-ray 
spectra of IPs. These include emission from hot thin plasma and 
soft blackbody radiation (dominating the spectrum of soft IPs in the ROSAT band).
In the following different combinations of these models are compared with the 
observed EPIC spectra. In all models absorption partially covering the X-ray source
is applied and an iron fluorescence line is included.

\subsubsection{Emission from thin plasma with constant temperature}

A spectral model consisting of hot thermal bremsstrahlung 
suffering absorption by matter partially covering the X-ray source together 
with an iron line complex (a fluorescent line from nearly neutral iron 
and two lines from \FeXXV\ and \FeXXVI) as it was representing X-ray spectra from
classical IPs in the past gives no acceptable fit to the EPIC spectra of \rxj. 
We used the MEKAL model \citep{1985A&AS...62..197M} for thermal
plasma emission from XSPEC 11.0.1 to account for the 
bremsstrahlung component and the two lines from highly ionized Fe 
(model A, see Table~\ref{epic-fit}). This resulted in a reduced $\chi^2$ of 3.5 with a large excess 
flux below 0.8 keV. 

\subsubsection{Thin plasma emission plus blackbody component}

Improved fit results were obtained by adding a blackbody component to account
for the low-energy excess (model B), improving the reduced $\chi^2$ to 1.30.
In this model the low absorption component of the partial absorber was also applied to the 
blackbody emission. Around 1 keV the fit is still not acceptable, again showing excess flux.
This excess indicates iron L emission as is also seen in high resolution
spectra from the IP RE\,0751+14 = PQ\,Gem \citep{2001burwitz}. 

\subsubsection{Two-temperature thin plasma emission plus blackbody component}

For iron L emission a lower temperature of the emitting matter is required
and we added another MEKAL component (soft thermal \ktlmk\ component in contrast to the
hard thermal \kthmk\ component) to the low absorbed part of the 
model spectrum (model C). This gives an acceptable fit over the whole 
energy band with reduced $\chi^2$ of 1.26 for 2113 dof. 

\begin{table*}
\caption[]{Spectral fit results for models A to E}
\begin{tabular}{c|cc|ccccc|c|c|cc}
\hline\noalign{\smallskip}
\multicolumn{1}{c|}{~} &
\multicolumn{2}{c|}{high absorption} &
\multicolumn{5}{c|}{low absorption} &
\multicolumn{1}{c|}{temp.} &
\multicolumn{1}{c|}{metal} &
\multicolumn{1}{c}{$\chi^2$} &
\multicolumn{1}{c}{dof$^a$} \\ 
& \multicolumn{1}{c}{\nh} & \multicolumn{1}{c|}{\kthmk} &
  \multicolumn{1}{c}{\nh} & \multicolumn{1}{c}{\kthmk$^b$} &
                            \multicolumn{1}{c}{\ktbb} &
			    \multicolumn{1}{c}{\ktlmk} &
			    \multicolumn{1}{c|}{\eqw} & 
			    \multicolumn{1}{c|}{distr. $\alpha$} &
			    \multicolumn{1}{c|}{abund.} & & \\
& \multicolumn{1}{c}{[\ohcm{23}]} & \multicolumn{1}{c|}{[keV]} & 
  \multicolumn{1}{c}{[\ohcm{21}]} & \multicolumn{1}{c}{[keV]} &
                                    \multicolumn{1}{c}{[eV]} &
				    \multicolumn{1}{c}{[keV]} &
				    \multicolumn{1}{c|}{[eV]} & & & & \\
\noalign{\smallskip}\hline\noalign{\smallskip}
A & 1.53          & 10.0         & 0.1           & 10.0 & --       & --            & 286        & --            & 0.34                   & 7444 & 2117 \\
B & 1.35$\pm$0.11 & 13.7$\pm$0.8 & 1.23$\pm$0.06 & 13.7 & 95$\pm$2 & --            & 272$\pm$27 & --            & 0.50$\pm$0.05          & 2748 & 2115 \\
C & 1.21$\pm$0.11 & 14.3$\pm$0.8 & 1.42$\pm$0.07 & 14.3 & 89$\pm$2 & 1.03$\pm$0.07 & 282$\pm$29 & --            & 0.53$\pm$0.05          & 2657 & 2113 \\
D & 0.97$\pm$0.10 & 59$\pm$10$^c$& 1.39$\pm$0.05 & 59   & 88$\pm$2 & --            & 253$\pm$24 & 1.0$^d$       & 0.98$^{+0.21}_{-0.12}$ & 2552 & 2115 \\
E & 0.95$\pm$0.08 & 60$^{c,d}$       & 1.47$\pm$0.08 & 60   & 86$\pm$2 & --        & 258$\pm$26 & 0.90$\pm$0.05 & 0.90$^{+0.10}_{-0.06}$ & 2544 & 2115 \\
\noalign{\smallskip}
\hline
\end{tabular}

For model A no errors were determined because of the unacceptable fit.\\
$^a$ degrees of freedom \\
$^b$ parameter linked to temperature of high absorption MEKAL component\\
$^c$ maximum temperature T$_{\rm max}$ in multi-temperature models\\
$^d$ fixed parameter
\label{epic-fit}
\end{table*}

\subsubsection{Multi-temperature thin plasma emission plus blackbody component}

\begin{figure}
\resizebox{\hsize}{!}{\includegraphics[angle=-90,clip,bb=112 38 554 548]{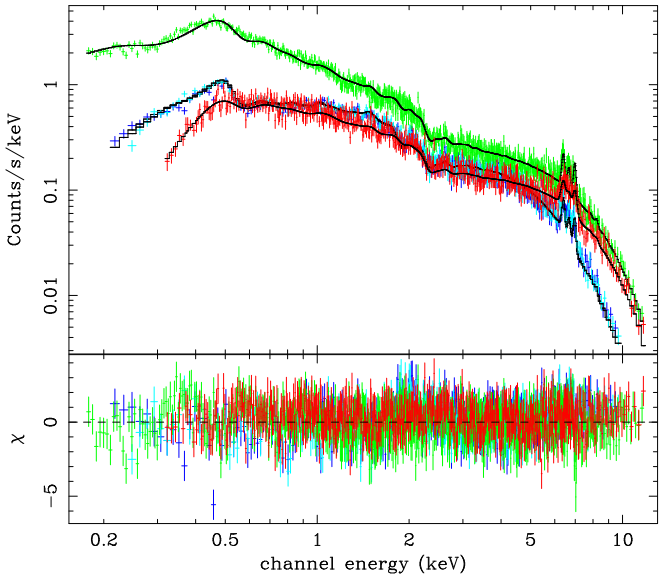}}
\caption{EPIC MOS (blue and light blue) and PN (single-pixel events green, 
         double-pixel events red) spectra of \rxj\ together with the best fit 
	 model E (see text). The bottom panel shows the residuals. Somewhat larger deviations near
	 0.5 keV and 2.2 keV may be caused by calibration uncertainties around 
	 instrumental absorption edges}
\label{epic-spectra}
\end{figure}

The co-existence of soft (1 keV) and hard (14 keV) thermal emission indicates
that we actually observe a wide range of temperatures. A multi-temperature 
model was used by 
\citet{1998MNRAS.298..737D} 
to fit ASCA and Ginga spectra from BY Cam. 
Such a model is available in the XSPEC package, based on the MEKAL 
model with emission measures which follow a power-law in temperature 
(the emission measure from temperature T is proportional to (T/T$_{\rm max}$)$^{\alpha}$).
To fit this model we replaced the two MEKAL components by the multi-temperature component,
suffering absorption again by matter partially covering the X-ray source.
As first approach we fixed the parameter $\alpha$ at 1.0 (model D in Table~\ref{epic-fit}).
The fit further improved to a reduced $\chi^2$ of 1.21 for 2115 dof (the $\chi^2$ decreased 
by $\sim$100 although the number of free parameters was reduced by 2). Alternatively
allowing $\alpha$ as free fit parameter the maximum temperature is not constrained
and was therefore fixed at 60 keV, the value derived with fixed $\alpha$.
This model (E in Table~\ref{epic-fit}) finally yields the best fit (reduced $\chi^2$ of 1.21)
although statistically not different from model D.
The spectra (PN singles+doubles, MOS1 and MOS2) are shown in Fig.~\ref{epic-spectra} 
together with the best fit model E.

As derived from the normalizations of the differently absorbed multi-temperature emission
components, the high column density matter covers the plasma by 47\% while
53\% are absorbed with a much reduced column density. 
The slope of the optical spectrum indicates that the column density of
the low absorption component, which corresponds to E(B-V) $\sim$ 0.25,
could entirely be of interstellar origin.
The flux determined from the spectra in the 0.2--10.0 keV band was 2.28\ergcm{-11},
2.34\ergcm{-11} and 2.30\ergcm{-11} for MOS1, MOS2 and PN respectively. 
PN single and double spectrum were forced to have the same normalization in the fit and
therefore yield the same flux. The fluxes derived from the three detectors 
agree remarkably well, it should be noted however that the used data does not
cover identical time intervals. The spectra are not corrected for point spread
function losses outside the source extraction radius. The used radius of
40\arcsec\ should encompass 93\% of the flux \citep{2000SPIE.4012..731A} with
little energy dependence.
The intrinsic 0.2--10.0 keV flux (with absorption set to 0) of the multi-temperature emission
component is 3.3\ergcm{-11} and the bolometric blackbody flux is 1.5\ergcm{-11}, yielding a 
ratio for blackbody to plasma emission of $\sim$46\%.
 
\begin{figure}
\resizebox{\hsize}{!}{\includegraphics[angle=-90,clip,bb=78 38 570 700]{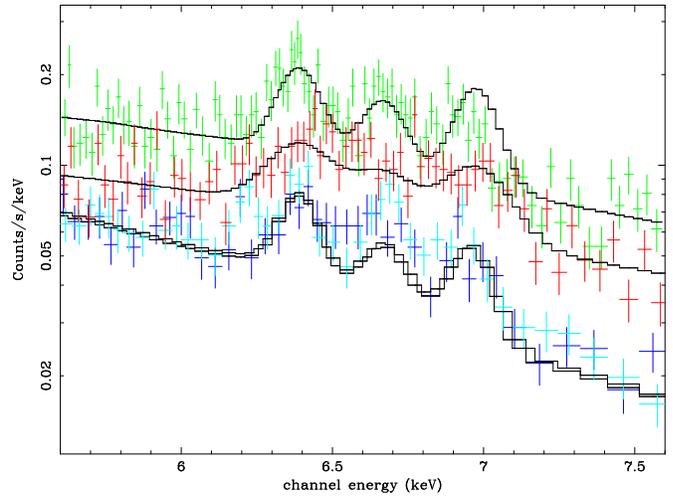}}
\caption{Enlarged part of Fig.~\ref{epic-spectra} showing the Fe line complex 
         in the EPIC spectra of \rxj. The colour coding is as in 
	 Fig.~\ref{epic-spectra}}
\label{epic-felines}
\end{figure}

The spectra around the Fe K emission line complex are shown enlarged in 
Fig.~\ref{epic-felines}. Iron fluorescence, \FeXXV\ and \FeXXVI\ lines 
are clearly resolved 
in the PN single-events and MOS spectra. The fluorescence line energy derived from 
the fits is 6.396$\pm$0.007 keV (models C-E). The equivalent width (\eqw) is listed in 
Table~\ref{epic-fit}. The \FeXXV\ and \FeXXVI\ line intensities require an 
under-abundance of iron in models A--C. We therefore allowed to vary the metal 
abundance in the MEKAL
components, but linked all metal abundances in all MEKAL components to one single
free parameter, which is mainly determined by the iron lines. The best fit values for 
the metal abundance varies considerably between models B--C with $\sim$0.5 solar and the 
multi-temperature models D and E where it is consistent with the solar value 
(Table~\ref{epic-fit}). For the equivalent hydrogen densities given in Table~\ref{epic-fit}
solar abundances in the absorbing matter are assumed. At least for the high absorption 
column, which is intrinsic to the \rxj\ system, reduced metal abundances by a factor of
two would increase the hydrogen column density by the same factor. 

There is indication for a stronger \FeXXV\ line than the model predicts. 
Producing more He-like iron emission relative to H-like iron emission
requires a lower temperature than the overall fit (which includes the
bremsstrahlung continuum) predicts. Fits to RXTE and BeppoSAX spectra of 
IPs extending to
energies up to nearly 100 keV yield reduced temperatures when a hard emission
component as arises from reflection on the white dwarf surface is
included in the spectral model \citep{2001A&A...377..499D}.

\subsection{Phase resolved spectral analysis}

Spectra were accumulated using events from spin phase 0.45--0.95 (spin maximum)
and 0.95--0.45 (spin minimum, see Fig.~\ref{pn-efold}). Subtracting the 
spin minimum from spin maximum spectrum shows an increasing count rate from 0.2
keV up to a maximum around 0.5 keV and then a decline down to 2 keV. This
suggests that mainly the two soft components (blackbody and 1 keV MEKAL in model C) 
change their intensity during the white dwarf spin revolution. In particular the
decrease of the difference spectrum towards low energies indicates that
absorption changes play a minor role.

\begin{figure}
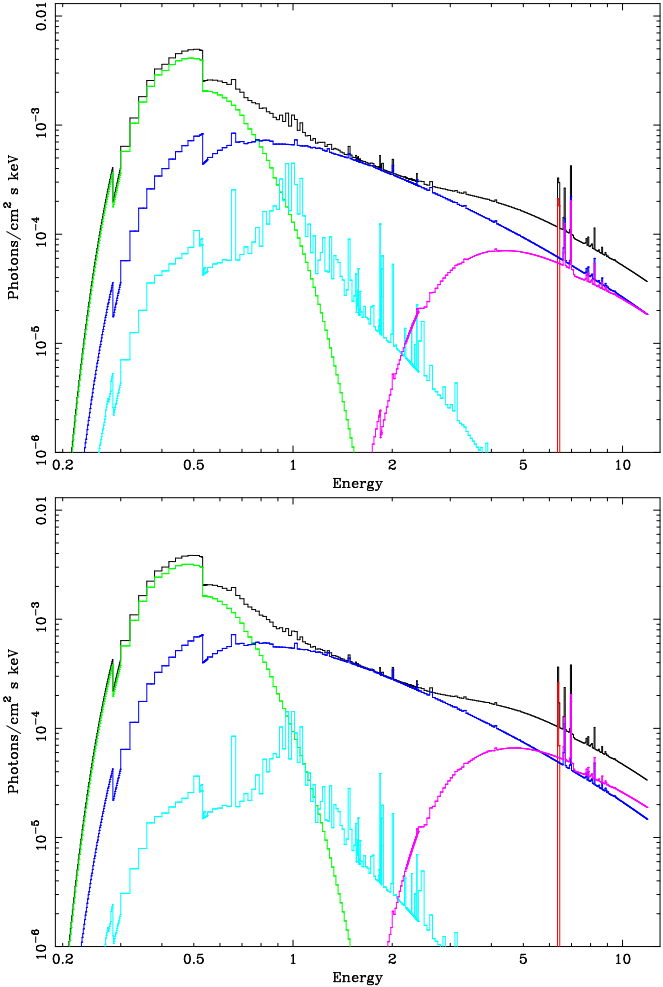

\resizebox{\hsize}{!}{\includegraphics[angle=-90,clip,bb=78 38 570 700]{h3394_model_phmax.ps}}
\resizebox{\hsize}{!}{\includegraphics[angle=-90,clip,bb=78 38 570 700]{h3394_model_phmin.ps}}
\caption{The spectral model components (model C described in the text) compared for spin
minimum (bottom) and maximum (top). Most of the intensity variations are caused by the 
soft thermal emission (light blue). 
This emission component shows an Fe-L line complex near 1 keV and is reduced
by a factor of $\sim$ 3 during spin minimum
        }
\label{epic-model}
\end{figure}

A fit to spin minimum and maximum spectra using models C and E shows that
temperatures and element abundances do not change. Therefore, these parameters
were fixed at the values derived from the spin averaged spectra. The
normalizations and column densities derived from the fits using models C and E
are compared for 
spin maximum and minimum spectra (for the case of PN) in Table~\ref{phase-fit}.
As suggested by the difference spectrum, the column density of the 
high absorption component
shows only a marginal change between spin minimum and maximum. The parameters for
absorption and temperature agree within the errors, although there is an
indication for some \nh\ decrease during spin maximum. 
Also the \nh\ values of the
low absorption components are consistent within errors. From the low absorption
components the soft thermal component (model C) shows the strongest 
intensity increase  
during spin maximum of at least a factor of 2.0 while the blackbody and the hard
thermal components vary at most by factors 1.7 and 1.3, respectively. This is 
illustrated in Fig.~\ref{epic-model} where the model components are drawn for
spin minimum and maximum. In model E the intensity increase of the soft
thermal emission is reflected in a decrease of the power-law index $\alpha$ of 
the temperature distribution.

\begin{table*}
\caption[]{Intensity variations in the spectral components (temperatures and element abundances
are fixed at values derived from spin averaged spectra)}
\begin{tabular}{c|cc|ccccc|c}
\hline\noalign{\smallskip}
\multicolumn{1}{c|}{Model/} &
\multicolumn{2}{c|}{high absorption} &
\multicolumn{5}{c|}{low absorption} &
\multicolumn{1}{c}{temp.} \\
\multicolumn{1}{c|}{Spin phase}  & \multicolumn{1}{c}{\nh} & 
\multicolumn{1}{c|}{\nhmk}       & \multicolumn{1}{c}{\nh} & 
\multicolumn{1}{c}{\nhmk}        & \multicolumn{1}{c}{\nbb} &
\multicolumn{1}{c}{\nlmk}        & \multicolumn{1}{c|}{\eqw} &
\multicolumn{1}{c}{distr. $\alpha$} \\
&                                  \multicolumn{1}{c}{[\ohcm{23}]} &
\multicolumn{1}{c|}{[10$^{-3}$]} & \multicolumn{1}{c}{[\ohcm{21}]} &
\multicolumn{1}{c}{[10$^{-3}$]}  & \multicolumn{1}{c}{[10$^{-4}$]} &
\multicolumn{1}{c}{[10$^{-4}$]}  & \multicolumn{1}{c|}{[eV]} & \\
\noalign{\smallskip}\hline\noalign{\smallskip}
C/Min. & 1.29$\pm$0.15 & 4.76$\pm$0.27 & 1.41$\pm$0.08 & 3.45$\pm$0.09 & 0.74$\pm$0.07 & 0.98$\pm$0.35 & 302$\pm$53 & --\\
C/Max. & 1.07$\pm$0.15 & 4.59$\pm$0.26 & 1.53$\pm$0.08 & 4.32$\pm$0.11 & 1.05$\pm$0.08 & 3.12$\pm$0.44 & 287$\pm$47 & --\\
\noalign{\smallskip}\hline\noalign{\smallskip}
E/Min. & 1.01$\pm$0.18 & 7.55$\pm$0.45 & 1.49$\pm$0.08 & 7.33$\pm$0.54 & 0.88$\pm$0.09 & -- & 277$\pm$53 & 0.95$\pm$0.06\\
E/Max. & 0.78$\pm$0.13 & 6.44$\pm$0.37 & 1.53$\pm$0.08 & 7.67$\pm$0.48 & 1.12$\pm$0.11 & -- & 265$\pm$41 & 0.76$\pm$0.04\\
\noalign{\smallskip}\hline
\end{tabular}
\label{phase-fit}
\end{table*}

The equivalent width of the Fe fluorescence line varies very little
between spin minimum and maximum. A small decrease during spin maximum
is insignificant given the large errors. If the fluorescence line would 
mainly originate on the white dwarf surface a correlation of its equivalent 
width with other emission from the surface may be expected. However, the 
blackbody emission increases during spin maximum suggesting that a large 
part of the line intensity is probably produced in the cold pre-shock 
matter of the accretion curtain which strongly absorbs the hard X-ray emission. 
In this case little changes in the equivalent width are expected as long as 
the line emitting volume remains visible to the observer.
In contrast to \rxj\ \citet{2001A&A...377..499D} 
detect variations in the equivalent width of the Fe fluorescence line
from RX\,J0028.8+5917 during a short interval in spin phase (of duration 0.2)
which we may miss due to our coarse phase resolution.

Together with the dip in hardness ratios around spin minimum
the spectral variations seen between spin minimum and maximum suggest that we are 
viewing the white dwarf nearly pole-on. If the soft thermal component originates
in the accretion curtain below the shock region, the variation in its intensity 
would then be caused mainly by the varying aspect angle under which the 
curtain is observed during the spin revolution 
and suggests beaming of the radiation preferentially out of the sides of the 
accretion curtains. This is consistent with the 
decrease in covering fraction of the high column density absorber
from 51\% at spin minimum to 46\% at maximum as calculated from the \nhmk\ values
for model E given in Table~\ref{phase-fit}.

\section{Discussion}

Our extensive optical and X-ray observations establish \rxj\ as an intermediate polar 
with rather unique properties. Optical spectroscopy and photometry indicate an orbital
period of 6.72 h or 9.37 h, probably together with GK Per (47.9 h), V\,1062\,Tauri 
(9.96 h) and AE Aqr (9.88 h) among the longest known from IPs. 
The optical spectrum is strikingly similar to that of \vcas\ \citep{2001A&A...374.1003B}
with broad absorption lines at \Hbet\ and higher order Balmer lines.
\citet{2001A&A...374.1003B} argue that these are produced in the non heated part of the white
dwarf atmosphere. Therefore, \rxj\ may be also a rare case where the three constituents 
of the system, hot white dwarf, accretion disc and mass donor star can be detected 
altogether at optical wavelength. This peculiarity opens prospects for an accurate 
determination of system parameters.

All X-ray observations of \rxj\ found the source at a similar X-ray flux level.
Using the spectral model parameters derived from the EPIC spectra expected
ROSAT PSPC and HRI count rates of 0.400 \ct\ and 0.146 \ct, respectively, 
were simulated.
This shows that the source was brightest during the Sept. 1996 observation,
a factor of 1.5 brighter than in Sept. 2000 when it was at the lowest observed 
intensity level.

To fit the X-ray spectra of \rxj\ as obtained by the EPIC instruments (0.15--12 keV)
a complex model is required. Thermal emission from a hot multi-temperature plasma 
in collisional ionization equilibrium together with absorption by matter partially
covering the hard X-ray source represents the spectrum above 1 keV. 
An approach with a two-temperature model suggests that we see a wide range of
temperatures covering at least 1 to 14 keV. High resolution spectra of magnetic
CVs are consistent with a continuous temperature distribution \citep{2001mauche,
2001burwitz} as is expected from the cooling matter in the post-shock accretion 
column. The fit of a model with a continuous temperature distribution results in a 
maximum plasma temperature of 60$\pm10$ keV, relatively unconstrained by the 
limited bandpass of the EPIC instruments. The temperature of the post-shock region
expected from the simple shock model 
kT = 3/8G$\mu$m$_{\rm{H}}$M$_{\rm{wd}}$/R$_{\rm{wd}}$ with the mean
molecular weight $\mu$ and the hydrogen mass m$_{\rm{H}}$, for a 1 M$_{\sun}$ white
dwarf with a radius of 0.008 R$_{\sun}$ 
\citep[for a mass radius relation see e.g.][]{2000A&A...353..970P} is $\sim$60 keV.
However this may be regarded only as upper limit because spectral models including
other hard spectral components result in decreased shock temperatures 
\citep{2000MNRAS.315..307B}. Fitting Ginga LAC spectra of V1223\,Sgr with and without
taking into account reflection of the hard bremsstrahlung from the white dwarf 
surface these authors found shock temperatures of 30$^{+12}_{-4}$ keV and 57 keV, 
respectively. Reducing the shock temperature to 30 keV translates to a white 
dwarf mass of $\sim$0.7 M$_{\sun}$. A more reliable shock temperature and
from this a better constrained estimate for the white dwarf mass requires the 
measurement of the hard X-ray spectrum at energies above 10 keV.

At energies below 1 keV a blackbody component with a temperature kT of 86 eV 
is needed to fit the EPIC spectra. The bolometric blackbody flux is 
$\sim$46\% of the intrinsic 0.2--10 keV hot plasma emission. This fraction is 
lower than that observed from AM\,Her systems and soft IPs, where the blackbody 
flux exceeds the bremsstrahlung emission. A correlation of the flux ratio
with the magnetic field strength at the main accreting pole suggests increasing 
suppression of bremsstrahlung in strong magnetic fields  
\citep{1995mcv..conf...99B}. 
The low flux ratio derived for \rxj\ is consistent with a magnetic field strength 
below 10 MG as generally anticipated for classical IPs.
The temperature of 86 eV is higher than that observed from AM\,Her systems and 
soft IPs which show typical values of 20--40 eV \citep[e.g.][]{2000A&A...358..177M} 
and 50-60 eV \citep{1994MNRAS.271..372D,1994A&A...291..171H}, respectively.
Models for the accretion process onto magnetic white dwarfs predict that the effective
temperature decreases with increasing size of the area covered by accretion on the 
white dwarf \citep{1990MNRAS.247..214K}. The lack of soft X-ray emission from IPs
supported this view and suggested temperatures too low for soft X-ray emission to be
detected. The case of \rxj\ and in this sense also the soft IPs show an opposite
behaviour: the temperature increases from AM\,Her systems via soft IPs to \rxj, i.e.
from high magnetic field strength (small accretion area) to low magnetic field strength
(large accretion area). This suggests that either \rxj\ is a very peculiar case or that
the temperature characterizing the soft emission is determined mainly by other accretion
parameters like the mass accretion rate or mass and radius of the white dwarf.
\citet{1999mcv..work....1H} also discussed that blobby accretion im AM\,Her systems
may lead to a larger footprint on the white dwarf, while in IPs accretion of more
homogenous matter from the accretion disc probably results in a narrow radial confinement
of the accretion curtain, compensating for the larger range in azimuth.

For an absorbing column density of \ohcm{23} the expected equivalent width of the
Fe fluorescent line should be of the order of $\sim$100 eV \citep{1985SSRv...40..317I}.
It is however not clear if the spherical geometry of the absorbing matter around 
the early type star in a high mass X-ray binary as used by \citet{1985SSRv...40..317I} 
can be transfered to the situation
we have in magnetic CVs with a more cylindrical geometry of the accretion stream.
From a tall cylinder with the line of sight along the cylinder axis high absorption
is expected but the cylinder contains relatively little matter to produce the 
fluorescence line. In a flat cylinder the absorption is low but the fluorescence
line stronger. In the accretion curtain scenario of IPs the geometry may be closer
to the latter case and indeed be the main reason for the high equivalent width of the 
Fe fluorescence line observed in \rxj\ and other IPs. To estimate the Fe line 
contribution from reflection by cold matter on the white dwarf surface a more 
detailed modeling of the Fe line production in the accreting matter is required
and arguments based on the results of \citet{1985SSRv...40..317I} sometimes used 
in the literature should be taken with caution.

The spectral analysis of \rxj\ has shown a strong model dependence of the derived metal 
abundances. While one- or two-temperature plasma emission models result in abundances of 
50\% solar, the models with continuous temperature distribution yield values compatible 
with solar elemental abundances. This is likely caused by insufficient spectral resolution 
in particular at low energies which does not allow to properly disentangle line 
from continuum emission. To constrain elemental abundances of the accreted matter
high resolution spectra obtained from grating spectrometers are required and values derived
from broad-band, medium resolution spectra should be treated with caution. 

\section{Conclusions}

The follow-up observations of \rxj\ at optical and X-ray wavelength have revealed a new
intermediate polar with interesting properties. The pulsations with a period of 693 s
most likely reflect the rotation period of the white dwarf and the optical data suggest
a long orbital period of more than 6 hours. Optical spectra are consistent with a late
type companion of spectral type K and the presence of broad absorption features underneath 
the Balmer emission lines could be the signature of the white dwarf photosphere, similar to 
the classical intermediate polar \vcas. The improved spectral resolution of the
XMM-Newton EPIC instruments together with the large collecting area of the X-ray mirrors
allowed a detailed study of the broad band 0.1-12 keV spectrum. The spectrum can be 
modeled by multi-temperature emission from hot plasma originating in the post-shock 
accretion region with a maximum temperature of $\sim$60 keV, an iron fluorescence line
with an equivalent width of $\sim$260 eV and a 86 eV blackbody component. About 47\% of 
the hard emission is absorbed by matter partially covering the X-ray source with a 
column density of \ohcm{23} while the remaining part and the blackbody emission are 
attenuated by a reduced column density of 1.5\hcm{21}. Intensity variations of 
individual spectral components between spin minimum and maximum suggest that the spin
modulation is mainly caused by viewing effects and only little changes in absorption.
The X-ray spectrum of \rxj\ and in particular the detection of soft blackbody emission
make this IP an interesting case between soft IPs and classical hard IPs.

\begin{acknowledgements}
The ROSAT and XMM-Newton projects are supported by the German Bundesministerium f\"ur
Bildung und For\-schung / Deutsches Zentrum f\"ur Luft- und Raumfahrt (BMBF/DLR), 
the Max-Planck-Gesell\-schaft and the Heidenhain-Stiftung.
\end{acknowledgements}

\bibliographystyle{apj}
\bibliography{cv,general,myrefereed,myunrefereed,mytechnical}

\end{document}